\documentclass[preprint,showpacs,preprintnumbers,amsmath,amssymb]{revtex4}
\usepackage{graphicx, epsfig}
\begin{document}

\title{Effect of Oxygen Content on  Magnetic Properties of Layered Cobaltites PrBaCo$_{2}$O$_{5+\delta}$}
\author{Shraddha Ganorkar, K. R. Priolkar, P. R. Sarode}
\address{Department of Physics, Goa University, Taleigao Plateau, Goa 403 206 India}
\author{A. Banerjee}
\address{UGC-DAE Consortium for Scientific Research, University Campus, Khandwa Road, Indore 452 001 India}

\date{\today}

\begin{abstract}
The effect of oxygen content on the magnetic properties of the layered perovskites has been investigated. The samples, PrBaCo$_{2}$O$_{5+\delta}$  (0.35 $\leq \delta \leq$ 0.80) were prepared by sol-gel method and characterized by X-ray diffraction  and DC magnetization. A detailed magnetic phase diagram for PrBaCo$_{2}$O$_{5+\delta}$ is presented.  It is found that unlike in the case of heavier rare-earths, ferromagnetic interactions are present at all temperatures below Tc and even in the antiferromagnetically ordered phase. Moreover, in compounds with lower oxygen content, short range ferromagnetic interactions are present even above Tc.  This dependence of magnetic properties on oxygen content in these layered perovskites has been linked to the changes in polyhedra around the Co ions.  
\end{abstract}


\maketitle

\section{Introduction}
RBaCo$_{2}$O$_{5+\delta}$, (R = rare earth elements; 0$ \leq \delta \leq $ 1) type double perovskites have attracted considerable attention due to considerably high magnetoresistance associated with insulator to metal transition. A strong overlap of the unfilled, and therefore magnetic, \textit{3d} electron orbitals with oxygen \textit{2p} orbitals in these compounds makes them display strong correlation between crystallographic, magnetic and transport properties. The oxygen content can also be tailored from $\delta$ = 0 to $\delta$ = 1 by appropriate heat treatment in different atmospheres \cite{mai}. The widely accessible oxygen nonstoichiometry and a strong tendency of oxygen vacancies to order in 112-type cobaltites, makes structural, transport and magnetic properties depend not only upon oxygen content itself, but also on the synthesis history \cite{str}.

The role of oxygen content is crucial in this family. Both, the mean valence of Co ions and the oxygen coordination around Co are controlled by it. The strength of the crystal field which determines the spin state of Co ions depends on the type of coordination cage surrounding Co ions. Furthermore, according to Goodenough-Kanamori rules the superexchange magnetic interactions are determined by the orbital occupancy of the outermost d-electrons.  Consequently the magnetic and transport properties of these compounds are strongly influenced by the oxygen content and the environment of Co ions. Another factor that is reported to play a role in magnetic properties of these double perovskites is the ordering of oxygen vacancies. The vacancy ordering is dependent on synthesis history \cite{str}. It is reported that oxygen disorder favours ferromagnetic or a canted antiferromagnetic ground state. \cite{str,fron,fron1}.  

The RBaCo$_{2}$O$_{5.5}$, in particular displays a variety of interesting phenomena including metal insulator transition \cite{mai,mar}, spin state transition \cite{mor,res},  charge ordering \cite{vog,sua} and giant magnetoresistance \cite{sua,tro}. The crystal structure consists of layers of CoO$_{2}$ - BaO - RO$_{0.5}$ - CoO$_{2}$ stacked along the \textit{c}-axis of an orthorhombic lattice. Due to an alternation of CoO$_{5}$ square pyramids and CoO$_{6}$ octahedra along the $\it b$ axis leads to its doubling. The CoO$_6$ octahedra and CoO$_5$ pyramids in these double perovskites are found to be heavily distorted \cite{kha,pom} which leads to presence of variety of Co$^{3+}$ spin states \cite{wu,zhi} as a function of crystallographic environment and temperature.

Furthermore, the magnetic properties are quite complex and the compounds display both ferromagnetic and antiferromagnetic transitions \cite{fron,cfr,lue}. Just below $T_{MI}$ the compounds undergo a paramagnetic (PM) to ferro(ferri)magnetic (FM) transition followed by a FM to antiferromagnetic (AFM1) transition which is accompanied by an onset of strong anisotropic magneto-resistive effects. The is also an AFM1 to second antiferromagnetic (AMF2) phase transition. The mechanism of such magnetic transformations at low temperatures is still not properly understood. It is also not clear why subtle changes in oxygen content should cause drastic changes in magnetic properties \cite{kim,bur}. On the basis of neutron diffraction \cite{fron,pla,fau,soda} and macroscopic measurements \cite{tas}, various contradicting magnetic structures, different spin states of Co$^{3+}$ ions as well as spin state ordering (SSO) have been proposed.  Muon-spin relaxation ($\mu$SR) studies
reveal that irrespective of the rare earth ion, a homogeneous FM phase with ferrimagnetic SSO of intermediate spin (IS)and high spin (HS) states develops through two first order phase transitions into phase separated AFM1 and AFM2 phases with different types of antiferromagnetic SSO \cite{lue}. Density functional calculations \cite{wu,wuh} and resonant photoemission studies \cite{fla} suggest that there is a strong hybridization between O-$2p$ and Co-$3d$ orbitals with a narrow charge transfer gap near Fermi level. With increasing temperature, the pd$\sigma$ hybridized hole in the O-$2p$ valence band suffers a gradual delocalization leading to successive magnetic transitions and spin reorientations.

In this work we focus our attention on the role of oxygen content on magnetic properties of these double perovskites. Oxygen content not only affects the valence of Co ions but also the local coordination around it. Such studies are limited to compounds with smaller rare-earth ions like Sm, Eu and Gd based cobalt double perovskites \cite{task,motin}. Here the ground state is an antiferromagnetic insulating phase for $\delta \approx 0.5$. Nanoscopic phase separation exists for lower values of $\delta$ while ferrimagnetic order is prevalent for higer values of $\delta$. A larger rare-earth ion like Pr, will affect the strength of the crystal field experienced by Co ions and therefore influence the magnetic properties. An understanding of magnetic properties is important from the point of view of both basic physics and probable applications.  Therefore we report here a detailed study of magnetic properties of PrBaCo$_{2}$O$_{5+\delta}$, (0.35 $\leq \delta \leq$ 0.80) prepared under very similar synthesis conditions.  

\section{Experimental}
The polycrystalline  samples of PrBaCo$_{2}$O$_{5+\delta}$ were prepared by sol-gel method. Stoichiometric amounts of  Pr$_{6}$O$_{11}$, BaCO$_{3}$ and Co$_{2}$(NO$_{2}$).6H$_{2}$O were dissolved in nitric acid. Citric acid was added to the above solution as a complexing agent. The solution was then heated at 353K to form a gel which was subsequently dried at 433K to remove the solvent. This precursor was ground, pelletized and heated at 1073 K for 4 hours followed by annealing at 1323 K for 24 hours and slow cooling at the rate of 1$^\circ$/min to room temperature to get the required sample. This as synthesized sample had an oxygen stoichiometry of $ \delta $ = 0.80 as determined by iodometric titration analysis. For iodometric titration, 15mg of sample and solid KI were dissolved in 1N HCl, and quickly titrated against 0.01N sodium thiosulphate using starch as an indicator. Inert atmosphere was maintained by adding sodium carbonate during titration. The oxygen content was calculated from the amount of sodium thiosulpahte consumed to get end point. Atleast three to five titrations were perfomed on each sample and the maximum probable error in the oxygen content was found to be less than $\pm$ 0.01. Oxygen content of each compound in the series was varied by individually annealing the samples in argon atmosphere at different temperatures. In particular, the samples with $\delta$ = 0.67, 0.58, 0.5 and 0.43 were annealed respectively at 673K, 873K, 1073K and 1173K for 36 hours  while the sample with $\delta$  = 0.35 was annealed twice at 1073K for 36 hours. Iodometric titration was again used to determine the oxygen content of these samples. The samples were deemed to be phase pure, as X-ray diffraction (XRD) data recorded using Rigaku D-Max IIC X-ray diffractometer  in the  range of $ 20^\circ \le 2\theta \le 80^\circ$ using Cu K$_\alpha$ radiation showed no impurity reflections. The diffraction patterns were Rietveld refined using FULLPROF suite and structural parameters were obtained. Magnetization measurements were carried out as a function of temperature and magnetic field using a Quantum Design SQUID magnetometer and VSM at different applied field values of 100, 1000 and 10000 Oe, in the temperature range of 10 K to 300 K. The sample was initially cooled from room temperature to the lowest temperature (10K) in zero applied field. Magnetization was recorded while warming under an applied field (zero-field cooled (ZFC)) and subsequent cooling field-cooled cooling (FCC) and warming (FCW) cycles. The isothermal magnetization M(H) curves were recorded at various temperatures in the field range of $\pm$ 7 Tesla and $\pm$ 14 Tesla.

\section{Results}
\begin{figure}
\centering
\includegraphics[width=\columnwidth]{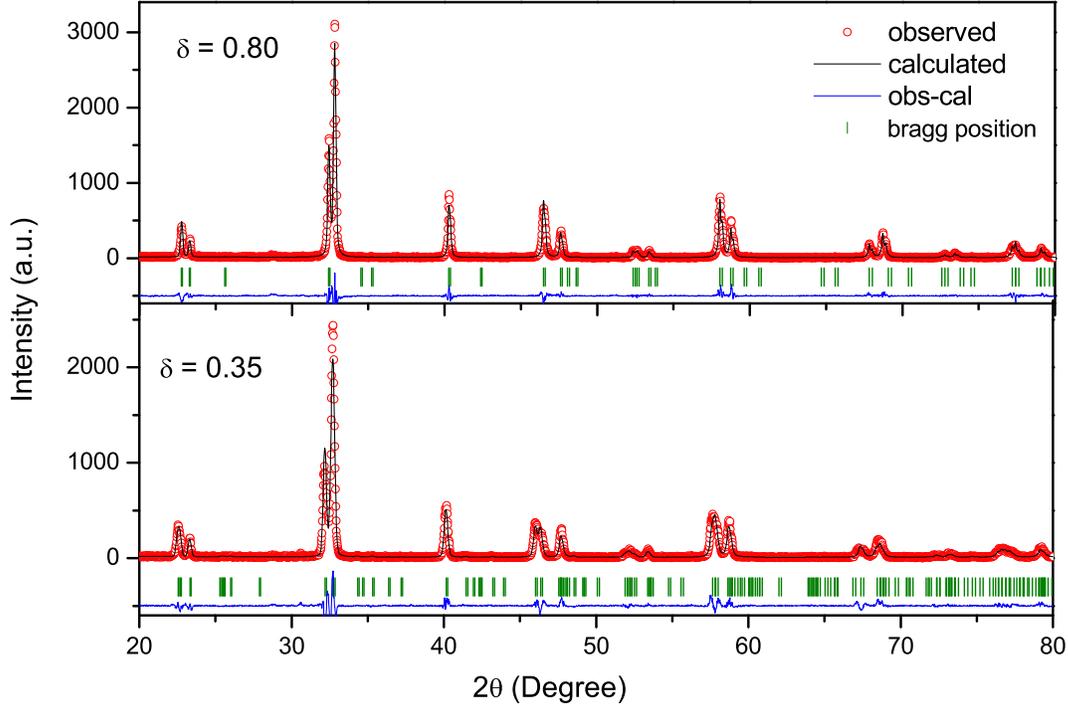}
\caption{\label{xrd} Rietveld refinement  XRD patterns of PrBaCo$_{2} $O$ _{5+\delta}$, $\delta$ = 0.80 and 0.35 . Circles represent experimental data, continuous line through the data is the fitted curve and the difference pattern is shown at the bottom as solid line.}
\end{figure}

\begin{figure}
\centering
\includegraphics[width=\columnwidth]{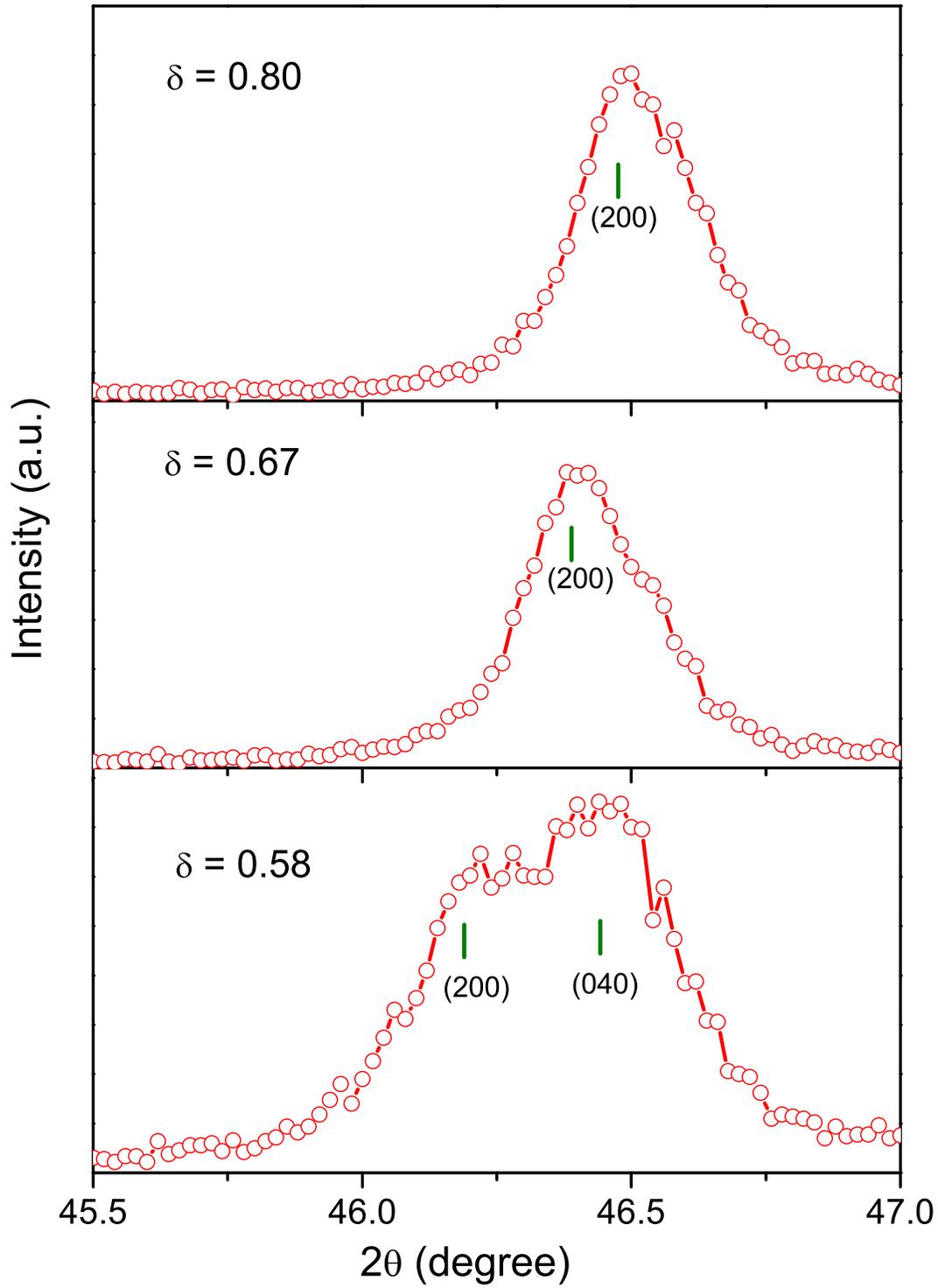}
\caption{\label{46xrd} The (200) Bragg reflection and its splitting to (200) (040) upon tetragonal to orthorhombic transition in PrBaCo$_2$O${5+\delta}$.}
\end{figure}

\begin{figure}
\centering
\includegraphics[width=\columnwidth]{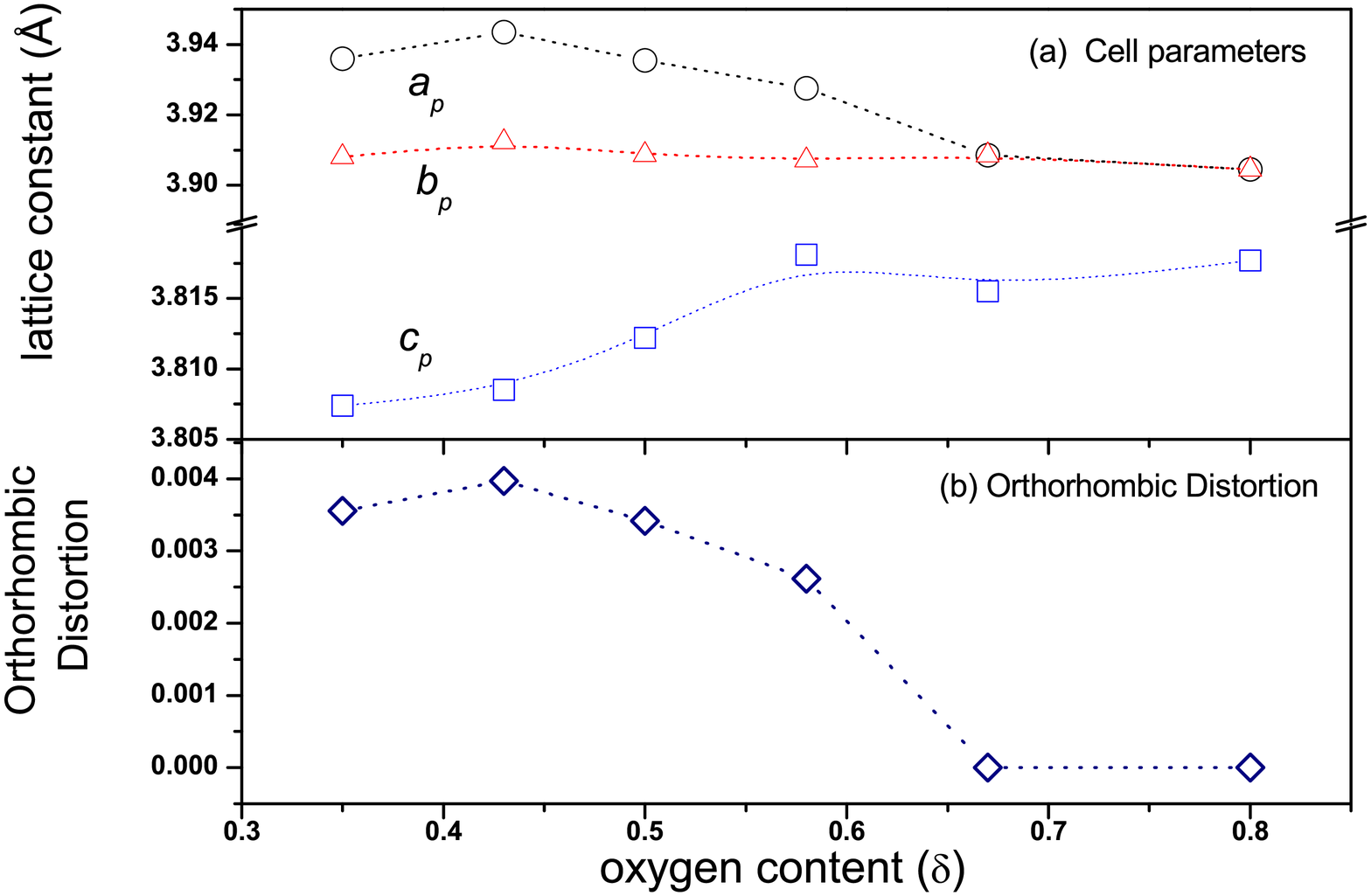}
\caption{\label{cell} Variation of cell parameters (a) and orthorhombic distortion (b) in PrBaCo$ _{2} $O$ _{5+\delta} $ as a function of oxygen content ($ \delta $). }
\end{figure}

The structure of Co based double perovskites is known to depend on the oxygen content $5+\delta$.  In general there are three structural regimes reported \cite{str}. The hole doped region ($0.7 \le \delta \le 1$) wherein the structure is tetragonal, the intermediate region ($0.3 \le \delta \le 0.7$) which is characterized by doubling of $b$ axis due to ordering of the Co ions and orthorhombic structure and the electron doped region ($0 \le \delta \le 0.3$) with the compounds having tetragonal structure. The limits of structural transition are not strict and often depend on the rare-earth ion. With the change in oxygen content, the coordination geometry around Co ions also changes from octahedral ($\delta \approx 1$) to square pyramidal($\delta \approx 0$). Fig. \ref{xrd} displays XRD pattern of as prepared PrBaCo$_2$O$_{5.80}$ sample and Ar annealed sample PrBaCo$_2$O$_{5.35}$. Rietveld refinement of XRD pattern of PrBaCo$_2$O$_{5.80}$ confirms the formation of a largely single phase sample with 112 type tetragonal unit cell belonging to $P4/mmm$ space group. A very minor impurity phase is detected with peaks around 2$\theta$ $\sim$ 29$^\circ$ and 31$^\circ$ which can be ascribed to unreacted Pr-oxides (Pr$_2$O$_3$, Pr$_6$O$_{11}$). These impurity peaks are seen only in the two end members and are not present in any of the intermediate compositions. The tetragonal structure preserves up to $\delta$ = 0.67. Samples with oxygen content $\le$ 5.58 crystallize in orthorhombic structure with $Pmmm$ space group and 122 type unit cell. This change in structural symmetry is reflected by the splitting of (200) peak at $2\theta\approx 46^\circ$ as can be seen in Fig. \ref{46xrd}.  Due to tetragonal to orthorhombic transition and subsequent doubling of unit cell along the $b$-axis, the (200) reflection splits into two distinct (200) and (040) Bragg peaks. A plot of variation of lattice parameters with oxygen content is presented in Fig. \ref{cell} From this plot it is observed that, with the decrease in oxygen content, the lattice parameters, $a$ and $b$ deviate away from each other upon the tetragonal to orthorhombic phase transition indicating an increase in orthorhombic distortion, $O_s = (a-b)/(a+b)$.  This distortion is found to be maximum for $\delta$ = 0.43 (Fig. \ref{cell}(b)). On the other hand, $c$ parameter remains nearly constant in the regions $0.8 \le \delta \le 0.6$ and $ \delta < 0.4$ and exhibits a decrease in for intermediate values of $\delta$. Such a trifurcation of lattice parameters is also seen in case of GdBaCo$_2$O$_{5+\delta}$ \cite{task}. It must be mentioned here that the regions indicated above are based on gross structural symmetry and local structural effect could play an important role in deciding the magnetic properties.

\begin{figure}
\centering
\includegraphics[width=\columnwidth]{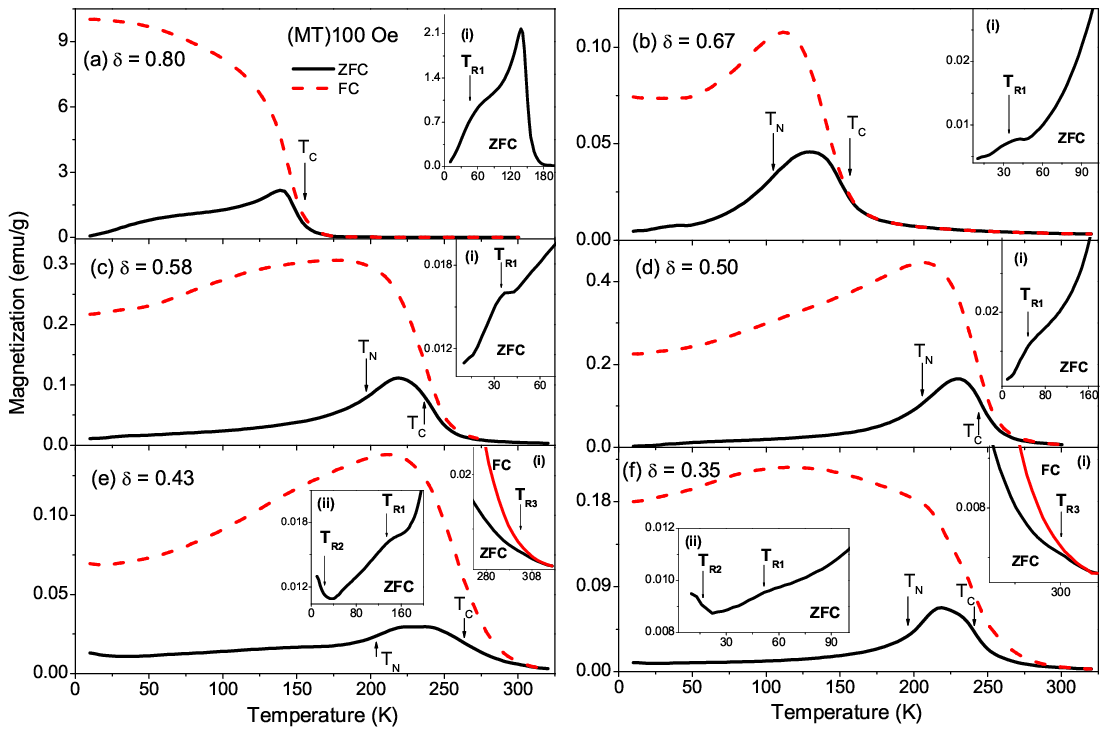}
\caption{\label{MT100} Magnetization as a function of temperature for PrBaCo$ _{2} $O$ _{5+\delta} $ compounds recorded in an applied field of 100 Oe. The solid lines denote magnetization recorded during ZFC cycle while the dashed curve represent magnetization during FC cycle. Insets show the details transition regions. }
\end{figure}

The temperature dependence of magnetization M(T) measured at 100 Oe is shown in Fig. \ref{MT100}. The ZFC and FC magnetization curves of all samples except $\delta$ = 0.80 show PM to FM transition followed by a decrease in magnetization which, based on neutron diffraction results \cite{fron}, is ascribed to a FM to AFM transition. The compound with $\delta$ = 0.80 exhibits a PM to FM transition at T$_C$ = 148 K. There is however, a large difference between ZFC and FC magnetization curves of this compound. Studies on oxygen rich PrBaCo$_2$O$_{5+\delta}$ indicate that compounds with $\delta$ = 0.9 and 0.85 order ferromagnetically while that with $\delta$ = 0.75 orders antiferromagnetically at 175 K which can be transformed to FM state under magnetic field \cite{gar}. Hence the difference between ZFC and FC magnetization curves for $\delta $ = 0.8 sample can be ascribed to the presence of competing magnetic interactions.

A closer examination of M(T) curves indicates presence of few more magnetic transitions at low temperatures. For $\delta$ = 0.80 apart from the PM to FM transition another transition is visible at  T$_{R1} \sim$ 50 K which has not been hitherto reported. This transition is also present in other compounds (see insets (i) in Fig. \ref{MT100} (a-d) and insets (ii) in e and f). The transition temperature T$_{R1}$ shows a tendency to increase with decreasing $\delta$ followed by a sharp rise to about 140 K for $\delta$ = 0.43. It may be noted that this particular compound has maximum orthorhombic distortion. The transition temperature decreases again to about 50 K for $\delta$ = 0.35. Muon spin relaxation ($\mu$SR) studies on NdBaCo$_{2}$O$_{5.5}$ attribute the transition around 50 K to a magnetic phase transition \cite{jarry}.

Furthermore, the magnetization behaviour of compounds with $\delta$ = 0.43 and 0.35 is even more complex. For example, in case of $\delta = 0.43$ a double hump structure is observed just below the PM to FM transition. Additionally a upturn in magnetization is observed at T$_{R2} \sim$ 17 K. This can be clearly seen in the inset (ii) of  Fig. \ref{MT100}(e) and (f) and could be ascribed to paramagnetic contribution of the rare-earth ion which is amplified in these two composition due to presence of Pr-oxide impurity phase seen in XRD. Furthermore, another new transition around 300 K = T$_{R3}$ (see inset (i) of Fig. \ref{MT100}(e) and (f)) is observed for $\delta$ = 0.43 and 0.35.

Second aspect that stands out is the low values of magnetization in compounds with $\delta < 0.8$. Different hypotheses have been considered to explain the low value of the magnetic moment. Firstly, it could be due to presence of low spin (S = 0) Co$^{3+}$ ions as majority species. However, optical studies on Eu based double perovskites discounts this possibility \cite{makh}. The resulting small total magnetization observed at low temperature could be due to ferrimagnetic ordering or mutually orientated moments of Co$^{2+/3+}$ ions in the nonequivalent pyramidal and octahedral sites. A third hypothesis is based on the coexistence of ferromagnetic and an antiferromagnetic phase or on segregation of a FM phase in the antiferromagnetic matrix in the form of ferromagnetic domains. This phase separation scenario is also observed for GdBaCo$_2$O$_{5+\delta}$ \cite{task}. In this case, the application of the magnetic field continuously transforms the antiferromagnetic state into ferromagnetic state.

\begin{figure}
\centering
\includegraphics[width=\columnwidth]{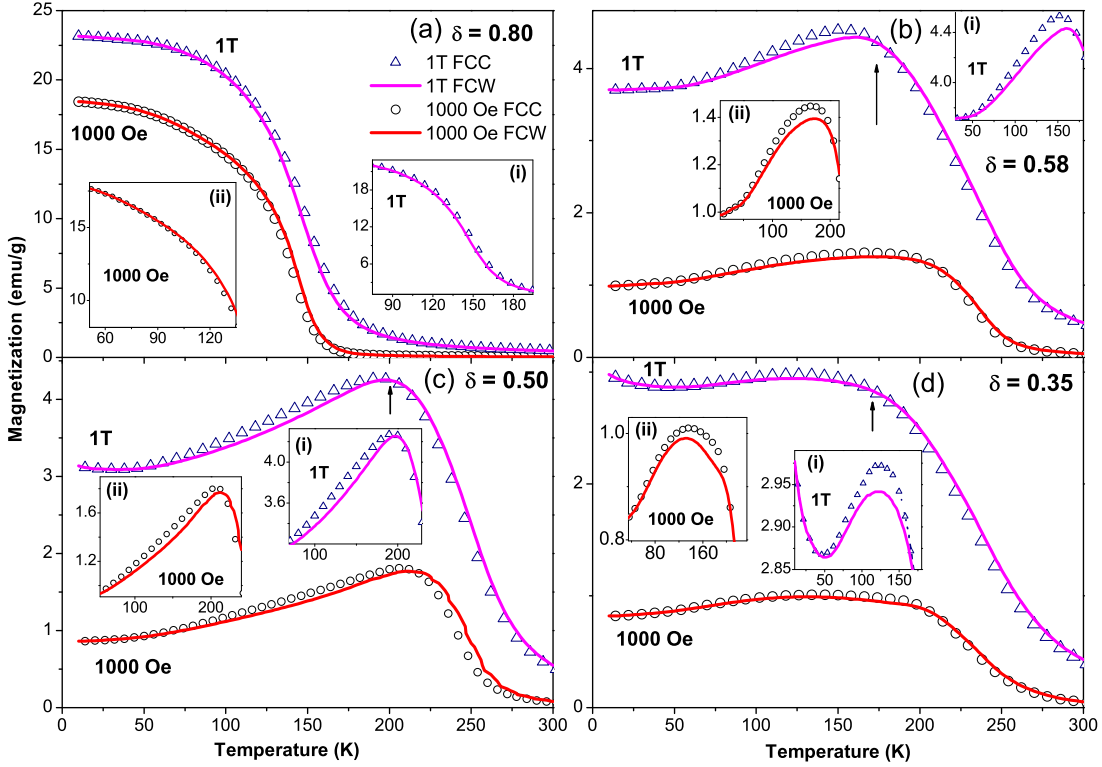}
\caption{\label{FCCFCW} Temperature dependance of magnetization recorded at 1000 Oe and 1 T for FCC and FCW cycles. The crossing FCC and FCW curves is presented more clearly in the insets.}
\end{figure}

To investigate the effect of magnetic field on the above magnetic transition temperatures and magnetization in general,  M(T) have been recorded at 1000 Oe and 1 T during FCC and FCW cycles for the samples with $\delta$ = 0.80, 0.58, 0.50 and 0.35.  A small amount of thermal hysteresis that increases with decreasing $\delta$ is still observed below T$ _{C} $ even in fields of 1 T confirming the presence of a complex magnetic ground state as was evident from magnetization curves at 100 Oe. Furthermore the absence of saturation is another evidence for the presence of competing FM and AFM interactions in these compounds. A closer look at Fig. \ref{FCCFCW} indicates a presence of a crossover between FCC and FCW curves (see insets (i) and (ii) of Fig. \ref{FCCFCW}(a-d)). At low temperatures the FCC curves lie below the FCW curves. As the temperature is increased, the FCC curves cross over and lie above the FCW curves. This cross over temperature is seen to be dependent on oxygen content. It is observed that this crossing temperature is maximum for $\delta$ = 0.50 (202 K), and lower for $\delta$ = 0.58 (176.4 K), 0.35 (174 K), while no such feature is observed for $\delta$ = 0.80. In the case of $\delta$ = 0.35, a similar crossover is also seen below 298 K indicating a presence of a short range magnetic transition.

\begin{figure}
\centering
\includegraphics[width=\columnwidth]{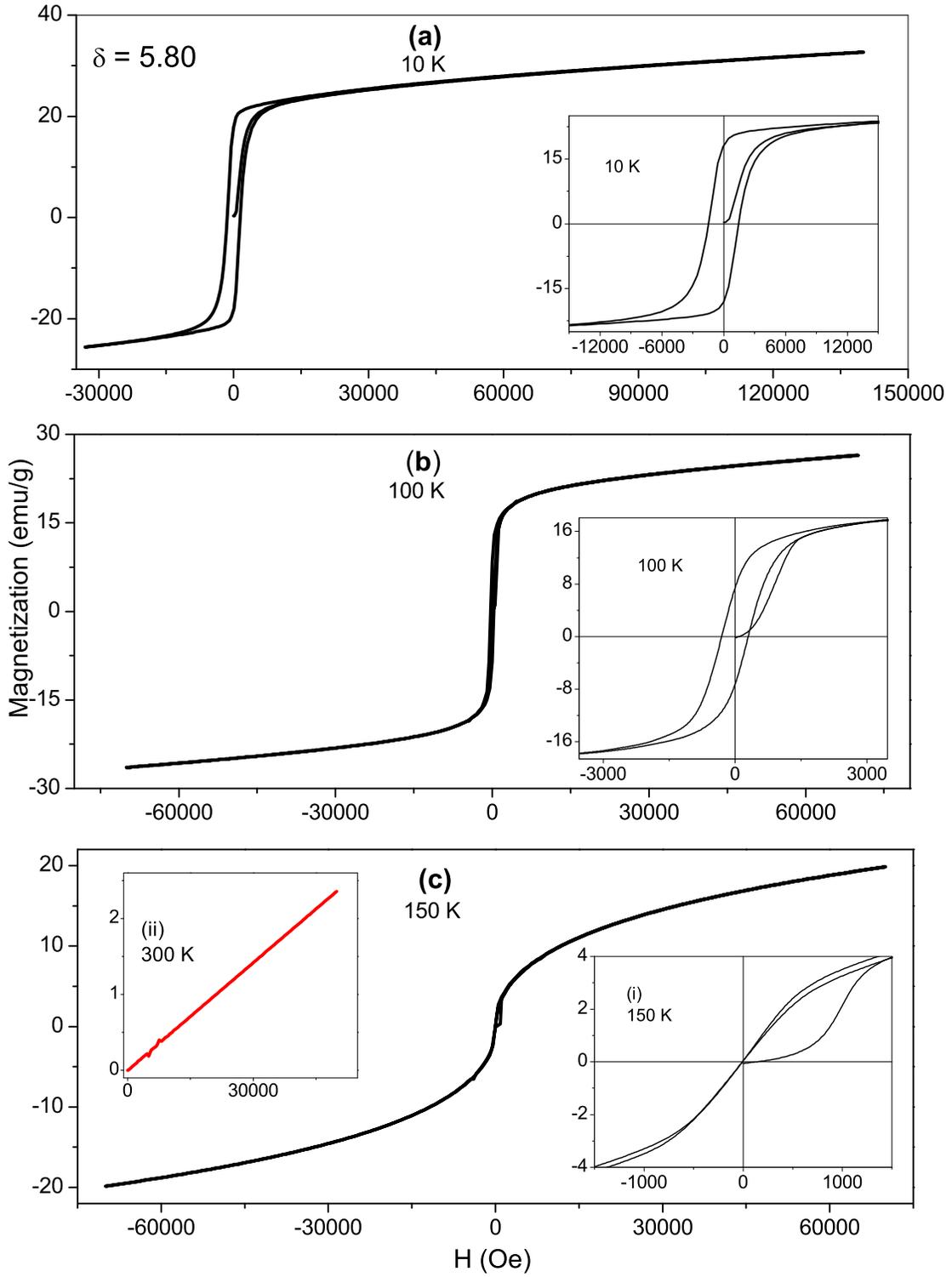}
\caption{\label{MH0.80} Hysteresis loops for PrBaCo$_{2}$O$_{5.80}$ recorded at (a) 10 K, (b) 100 K and (c) 150K. The same data on a magnified scale is shown in the insets. M v H data at 300K is presented in inset(ii).}
\end{figure}

The isothermal magnetization M(H) recorded at 10 K and in the vicinity of T$ _{C} $ and T$ _{N} $ for compounds having $\delta$ = 0.80, 0.58 and 0.50 compounds and at 250K and 300K for $\delta$ = 0.35. The M(H) for $\delta$ = 0.80 is presented in Fig. \ref{MH0.80}. A clear hysteresis loop signifying the presence of ferromagnetism is observed at 10 K. The magnetization however, does not saturate even up to 14 T indicating a presence of strong magnetic anisotropy or a competing AFM state. The hysteretic behaviour persists up to 150 K though the coercive field decreases with increasing temperature before vanishing at T$_C$ = 150 K (top left inset of Fig. \ref{MH0.80}(c)). M(H) curve at 300 K shown in the bottom left inset of Fig. \ref{MH0.80}(c) indicates the sample to be paramagnetic at this temperature. Another notable feature in the hysteresis curves is that the virgin curve lies out side the loop at 100K and 150K. This can be attributed to  the presence of two different Co ions in the matrix. Due to the presence of Co$ ^{3+}$ and Co$^{4+}$ ions there is an electronic phase separation. The ferromagnetism can be attributed to majority Co$^{3+}$ - O - Co$^{4+} $ superexchange interactions in agreement with Goodenough-Kanamori rule \cite{good}. While AFM can be due to minority Co$^{3+}$ - O - Co$^{3+} $ superexchange interaction. In the presence of magnetic field the contribution from FM interactions grows at the expense of AFM contribution
 leading to the initial magnetization curve to lie outside the main loop. 

\begin{figure}
\centering
\includegraphics[width=\columnwidth]{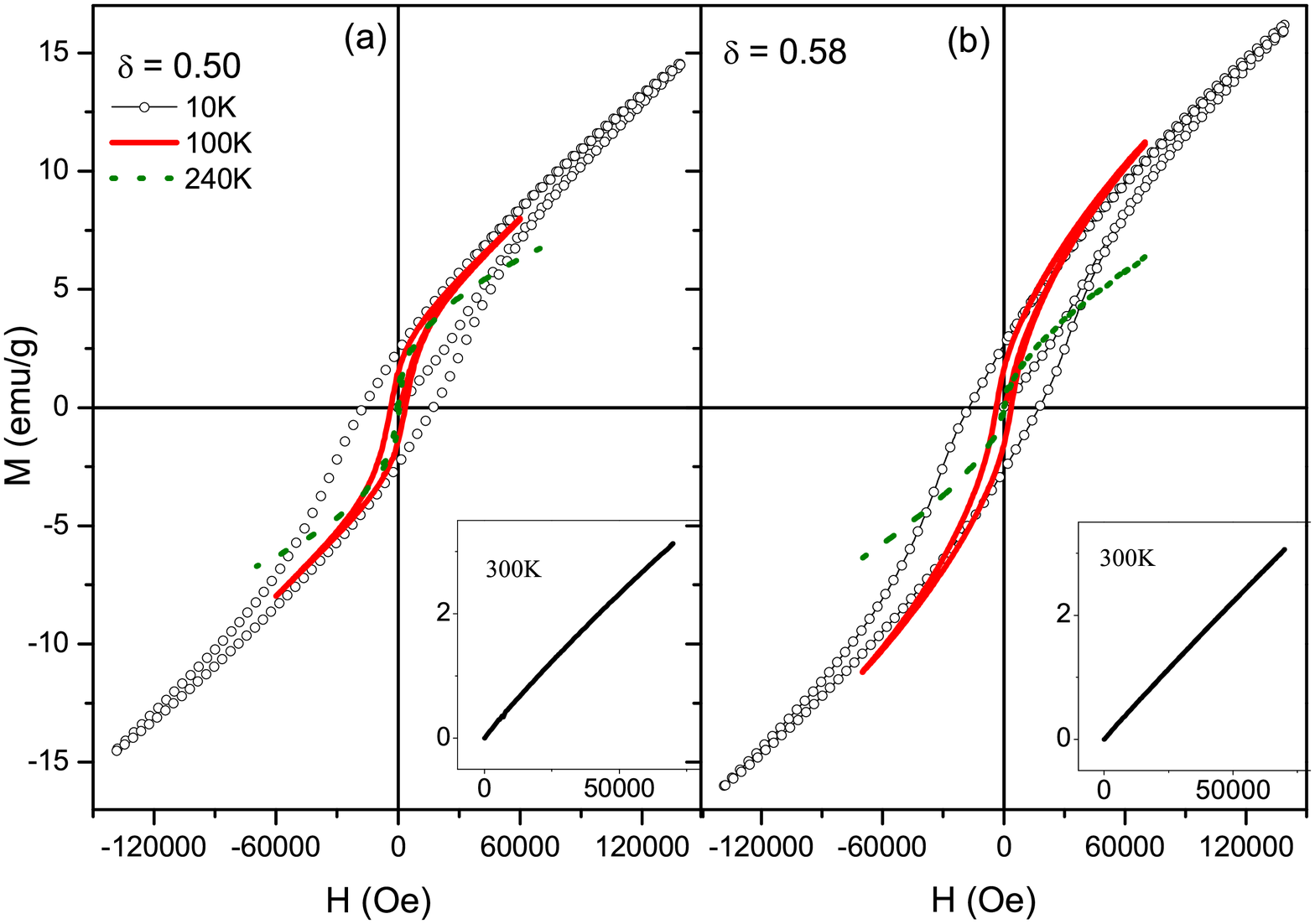}
\caption{\label{MH0.5_0.58} Isothermal magnetization curves of PrBaCo$ _{2} $O$ _{5+\delta}$.(a) $\delta = 0.50$ (b) $\delta = 0.58$ recorded at 10 K, 100 K, and 240 K. The initial magnetization curve at 300 K is shown in inset (a) and (b).}
\end{figure}

The M(H) curves for PrBaCo$_2$O$_{5.5}$ and PrBaCo$_2$O$_{5.58}$ are presented in Figs. \ref{MH0.5_0.58}. The behaviours of the two samples being very similar these are discussed together. The main point is that although the samples are reported to undergo AFM transition, ferromagnetic hysteresis loop is observed at all temperatures below T$_C$. Further, for the sample with $\delta$ = 0.5, isothermal magnetization measured at 300K (Inset in Fig. \ref{MH0.5_0.58}(a)) shows a slight curvature indicating presence of short range ferromagnetic interactions well above $T_C$. In the case of both samples, coercive field increases with decrease in temperature indicating strengthening of FM interactions. However, no saturation of magnetization is observed even at lowest temperature and maximum field of measurement in the case of these compounds. The continuous increase in magnetization with field could also be attributed to paramagnetic contribution of Pr ions. However, this contribution would have been reversible, in contrast to the obtained result. The hysteresis loop at 10K shows a clear irreversibility right up to fields = 14T. This indicates presence of competing ferro and antiferromagentic interactions in the two samples. Moreover, it is clear that the magnetization is induced by the applied field rather than being of spontaneous nature.

\begin{figure}
\centering
\includegraphics[width=\columnwidth]{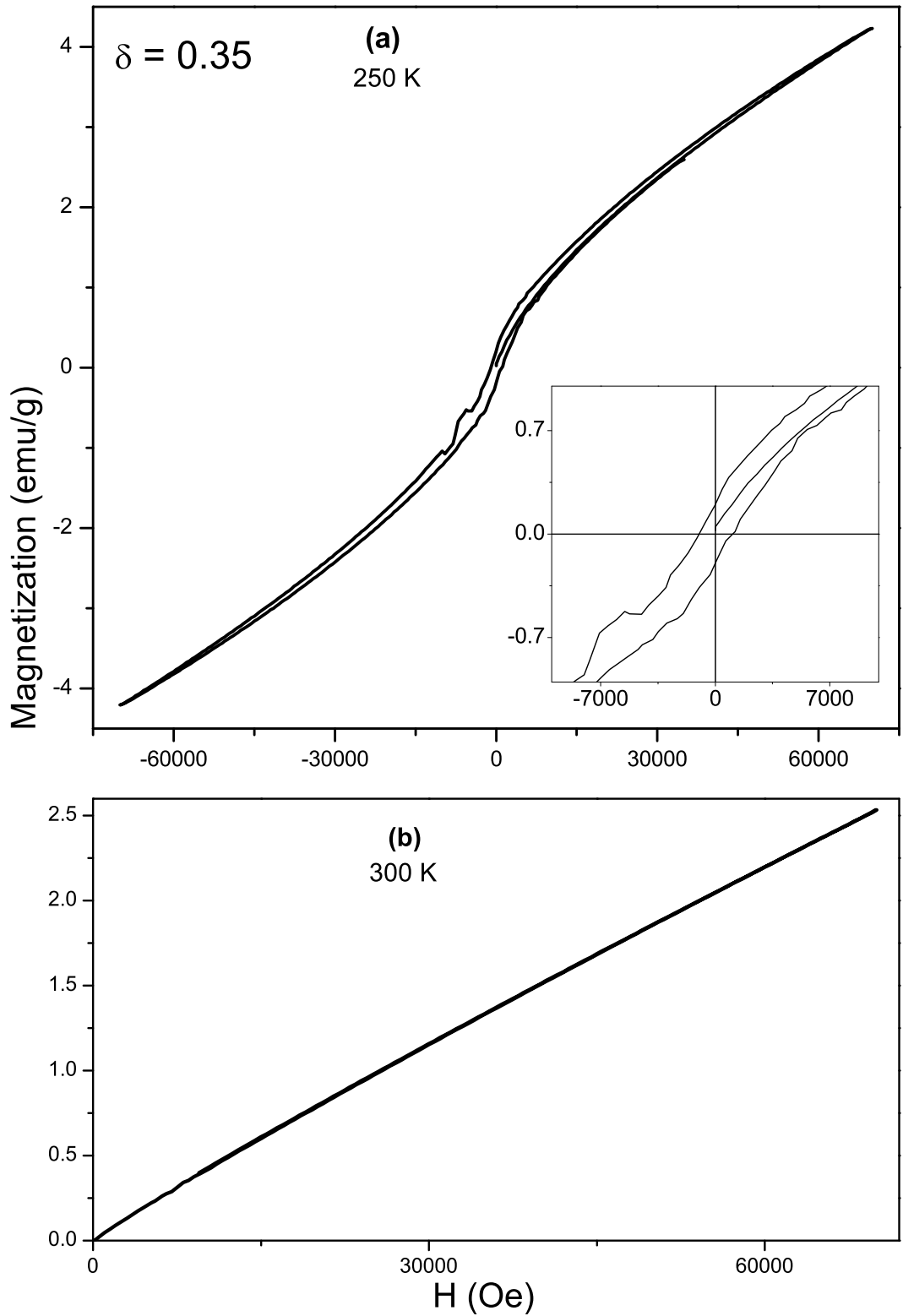}
\caption{\label{MH0.35} Magnetization as a function of magnetic field for PrBaCo$ _{2} $O$ _{5.35}$ recorded at (a) 250 K with the inset showing expanded view of the hysteresis loop and (b) the data recorded at 300 K.}
\end{figure}

In case of compound with low oxygen content, ($\delta$ = 0.35), M(H) curves at 300K and 250K are presented in Fig. \ref{MH0.35}. It may be noted that both temperatures are above its T$_C$.  The magnetization curve at 300 K exhibits a curvature towards the horizontal axis. Such a behaviour can be attributed to presence of short range ferromagnetic interactions. Furthermore, the M(H) curve at 250 K exhibits a narrow hysteresis loop (m$ _{0} \simeq $ 0.21 emu/g, H$ _{C} \simeq$ 856 Oe). This observation clearly indicates the transition T$_{R3}$ to be ferromagnetic in nature. This weak FM ordering could arise due to Co$^{2+} - O - Co^{3+}$ interactions.

\section{Discussion}
\begin{figure}
\centering
\includegraphics[width=\columnwidth]{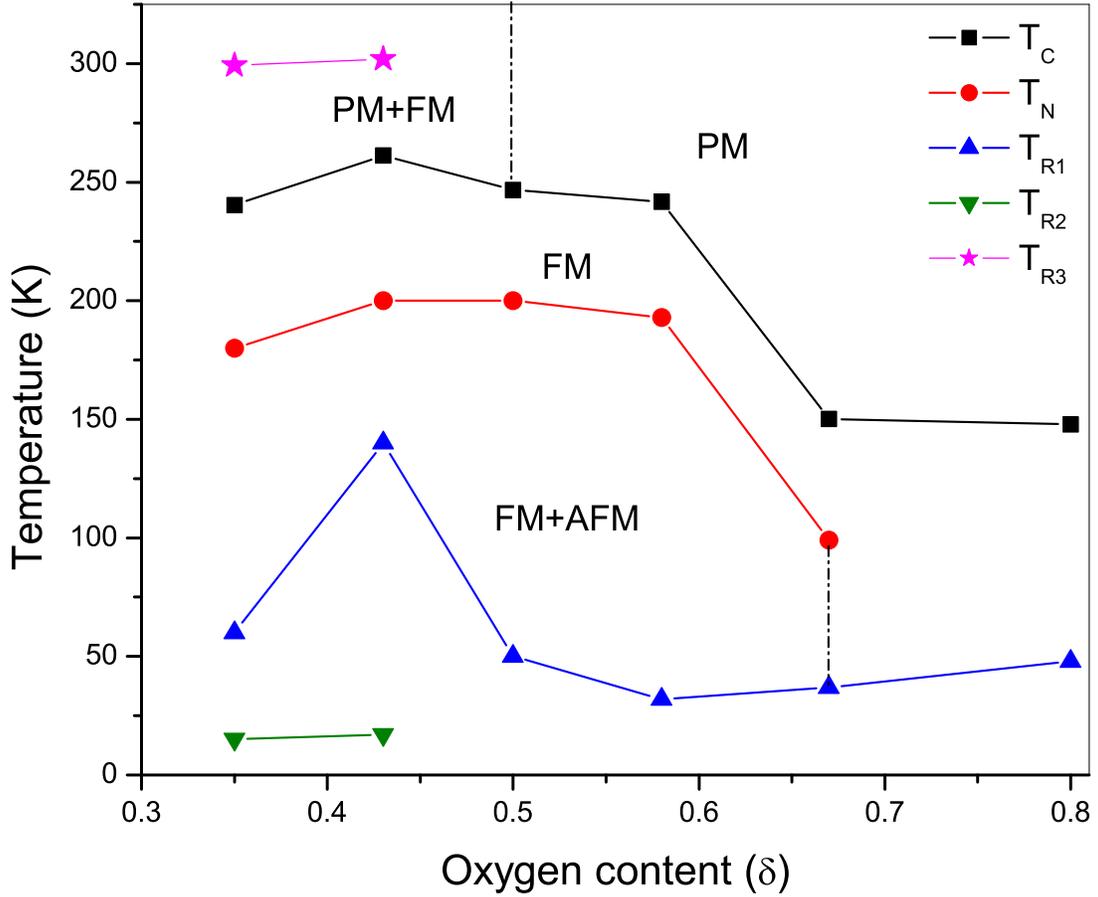}
\caption{\label{tcurve} Effect of oxygen content on various transition temperatures  estimated from 100 Oe ZFC magnetization data for  PrBaCo$_{2}$O$_{5+\delta}$.}
\end{figure}

It is quite evident from the above studies that $\delta$ has a great influence on magnetic transition temperatures. The variation of these transition temperatures as identified from the magnetization curves recorded at 100 Oe as a function of oxygen content ($\delta$) is depicted in Fig. \ref{tcurve}. It is quite evident that both T$_{C}$ and T$_{N}$ increase as $\delta$ decreases from 0.67 to 0.35. Though the magnetization of sample with $\delta$ = 0.8 does not show presence of an antiferromagnetic transition, the wide separation between ZFC and FC magnetization curves and the nature of magnetic hysteresis loops clearly show presence of competing AFM and FM interactions. It is also seen that all the compounds undergo another magnetic transition T$_{R1}$ at lower temperatures. This transition appears to be related to orthorhombic distortion. In addition, the compounds with $\delta \le 0.43 $ exhibit two new magnetic transitions, T$_{R2}$ and T$_{R3}$. Presence of hysteresis in M(H) loops of $\delta$ = 0.35 well above its T$_C$ tends to indicate that the transition at T$_{R3}$ is ferromagnetic. Based on these, the nature of magnetic order present in different temperature zones as a function of oxygen content is indicated in Fig. \ref{tcurve}. The magnetic phase diagram so constructed is quite different from that obtained for cobalt double perovskites containing heavier rare-earth ions like Eu and Sm \cite{motin}. Although the Pr containing double perovskites undergo more than one magnetic transition which include PM to FM and FM to AFM transitions, the pure antiferromagnetic phase found in perovskites with heavier rare-earths is absent in PrBaCo$_2$O$_{5+\delta}$. Here, ferromagnetism prevails and co-exists with antiferromagnetic phase right down to the lowest temperature studied. It appears that for compounds with lower values of $\delta$ ferromagnetic phase is present as a minor phase or domains embedded in an antiferromagnetic matrix, while in case of $\delta = 0.8$, the compound is ferromagnetic due to Co$^{3+}$ - O - Co$^{4+}$ superexchange interactions with a minor antiferromagnetic phase arising due to Co$^{3+}$ - O - Co$ ^{3+}$ interactions.

\begin{figure}
\centering
\includegraphics[width=\columnwidth]{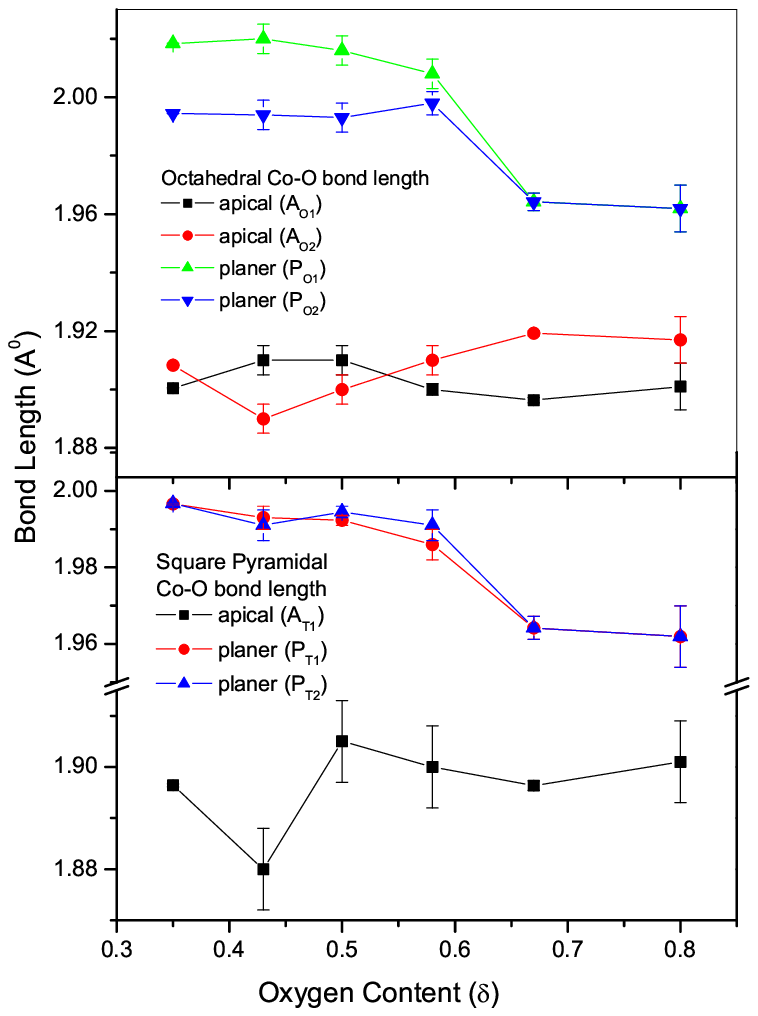}
\caption{\label{blength}Variation of Co-O bond lengths obtained from room temperature XRD data for octahedral and square pyramidal coordination of PrBaCo$ _{2} $O$ _{5+\delta} $ as a function of oxygen content $\delta$. }
\end{figure}

A plot of various Co-O bond distances obtained from Rietveld refinement of room temperature XRD patterns against oxygen content $\delta$ is presented in Fig. \ref{blength}. It can be seen that the Co-O planar bond distances belonging to both the polyhedra increase when the structure changes from tetragonal to orthorhombic. Further the variation of the apical bond distances indicate that as $\delta$ decreases the octahedra shrink from top and get elongated from bottom. An increase in planar Co-O bond length with structural transition indicates the tilting of polyhedra. Tilting of polyhedra with decrease in oxygen content results in increased  distortion of the CoO polyhedra. Distortion of polyhedra reflects in increased orthorhombic distortion with decreasing $\delta$. Increased distortions will also result in a decrease in Co - O - Co inter polyhedral bond angle and thereby favour antiferromagnetic ordering due to localization of charge carriers. However, the presence of a relatively larger rare-earth ion like Pr tends to minimize these Co-O polyhedral distortions.  Thus resulting in the ferromagnetic phase persisting right down to the lowest temperature. A detailed investigation of local structural distortions and their variation as a function of temperature will help in understanding the magnetic phase diagram.

\section{Conclusion}
In summary, there exists a transition from tetragonal to orthorhombic structure with decrease in oxygen content. All the samples have a complex magnetic ground state due to competing  ferromagnetic and antiferromagnetic interactions. Unlike in the case of heavier rare-earths where a pure antiferromagnetic ordering has been observed, the phase below T$_N$ in PrBaCo$_{2}$O$_{5+\delta}$ is a mixture of antiferromagnetic and ferromagnetic order. In addition to T$_C$ and T$_N$, all the compounds undergo another magnetic transition at T$_{R1}$. The temperature of this transition is highest for the compound with largest orthorhombic distortion indicating it to be linked to structural degrees of freedom. Compounds with lower oxygen content ($\delta \le 0.43$), undergo two more magnetic transitions - one around 15 K (T$_{R2}$) and the other at around 300 K (T$_{R3}$). Presence of magnetic hysteresis indicates the transition at T$_{R3}$ to be ferromagnetic in nature. The multiple magnetic transitions can be ascribed to the distortions of CoO polyhedra arising due to oxide ion deficiency.

\section*{Acknowledgements} Authors would like to acknowledge the financial assistance from Council for Scientific and Industrial Research, New Delhi under the project EMR-II/1099. SG acknowledges the travel assistance and local hospitality extended to her by Centre Director, UGC-DAE Consortium, Indore.

\end{document}